# Tracing Digital Footprints to Academic Articles: An Investigation of PeerJ Publication Referral Data

Xianwen Wang[1,*], Shenmeng Xu[2] and Zhichao Fang[1]

1. WISE Lab, Faculty of Humanities and Social Sciences, Dalian University of Technology, Dalian 116085, China

2. School of Information and Library Science, University of North Carolina at Chapel Hill, Chapel Hill, NC 27599-3360, USA

* Corresponding author.
Email address: xianwenwang@dlut.edu.cn; xwang.dlut@gmail.com
Website: http://xianwenwang.com

**Abstract:** In this study, we propose a novel way to explore the patterns of people's visits to academic articles. About 3.4 million links to referral source of visitors of 1432 papers published in the journal of *PeerJ* are collected and analyzed. We find that at least 57% visits are from external referral sources, among which General Search Engine, Social Network, and News & Blog are the top three categories of referrals. Academic Resource, including academic search engines and academic publishers' sites, is the fourth largest category of referral sources. In addition, our results show that Google contributes significantly the most in directing people to scholarly articles. This encompasses the usage of Google the search engine, Google Scholar the academic search engine, and diverse specific country domains of them. Focusing on similar disciplines to *PeerJ*'s publication scope, NCBI is the academic search engine on which people are the most frequently directed to *PeerJ*. Correlation analysis and regression analysis indicates that papers with more mentions are expected to have more visitors, and Facebook, Twitter and Reddit are the most commonly used social networking tools that refer people to *PeerJ*.

## 1. Introduction

In print publishing era, it was challenging to trace scholars' usage of journal articles in spite of subscription records and library catalog information. As the digitalization of the academic publishing industry and scholarly communication infrastructures, scholars are increasingly working digitally. These digital workflow encompasses diverse information acts including searching for scholarly information, saving and organizing the information, as well as the utilization of the information. Among these, the first step before the following information organizing, keeping and creating behaviors is to find the information. Before, no traces could be tracked when scholars read printed articles; however, nowadays digital footprints are created when users visit the webpage of academic articles. By exploring these digital footprints, we have the opportunities to better understand the patterns of scholars' visits to academic articles. Although these data are not generally provided by all publishers, *PeerJ*, an Open Access publisher launched in 2012, display the usage data for each of its articles on their webpages (see Figure 1). Are scholarly search engines used more frequently in searches for scholarly articles? How many article links in tweets are clicked on when browsed? How many people go to read the original scientific article when reading news mentioning them? Using PeerJ's data, our study presents a novel way to make use of digital footprints to explore these questions.

For decades, citations of scholarly publications have been regarded as a proxy of research impact; the advent of altmetrics (Priem, Taraborelli, Groth, & Neylon, 2010) has brought new opportunities to evaluate the impact of academic outputs in a broader range. Using mentions of scientific papers in social media, traditional news and blogs, as well as online reference management tools (Costas, Zahedi, & Wouters, 2014), etc., altmetrics provide the possibility to track the impact of newly published papers before they accumulate citations (Kwok, 2013). Many studies have explored the meaning (Bornmann, 2014; Galligan & Dyas-Correia, 2013; Haustein, Bowman, & Costas, 2015; Lin & Fenner, 2013), validity (Li, Thelwall, & Giustini, 2011; Thelwall, Haustein, Larivière, & Sugimoto, 2012), and utility of altmetrics (Cheung, 2013; Konkiel & Scherer, 2013; Piwowar, 2013; Piwowar & Priem, 2013). Unlike citation, which is believed to reflect scholarly impact, altmetrics is considered to be able to measure a broader societal impact of scholarly works (Bornmann, 2014). According to a survey conducted in 2014, correlations between tweets and citations are low, which implies that the impact metrics based on tweets are different from those based on citations (Haustein, Peters, Sugimoto, Thelwall, & Larivière, 2014).

Nowadays, as scholars are increasingly working digitally, we see more opportunities for investigation into their information behaviors. More and more academic publishers are providing Article Level Metrics (ALM) data to the public (Wang, Mao, Xu, & Zhang, 2014; Wang, Wang, & Xu, 2013; Wang et al., 2012). Article Level Metrics (ALM) data could be accessed from a number of providers like Altmetric LLP, ImpactStory, and Plum Analytics, etc. In addition, the detailed information about the traces as opposed to only the counts of the traces are also offered, which makes it feasible to investigate the context of the traces, thus enrich the meaning of the figures. For instance, by looking at news, blogs and tweets content that link back to one article, we could better understand what the impact is to the audience, why the article is impactful, and how the impact could be potentially utilized in the future.

Nevertheless, most previous altmetrics studies have been focusing on how to use digital traces of audience' behaviors to interpret the assumed impact; few studies have looked specifically into the actual behaviors and how impact is reflected by them accordingly. For instance, one article would have different impact on people who searched this article on Web of Science and those who clicked on a link to this article on a Reddit post. Although these two traces both appear to be "views", they are different from each other in essence. One of them is likely to be associated with the behavior of intended scholarly information seeking, while the other is more likely to be induced by serendipity luck in social media browsing. By looking at the types and counts of traces, we could barely know the behaviors and information seeking purposes underneath the traces; however, by looking at referral data, which indicate where the audience come from, we could get a better understanding of them.

Previous studies have investigated the diffusion of scientific papers on the web (Christozov & Toleva-Stoimenova, 2013; Herie & Martin, 2002; Priem & Costello, 2010; Wang, Mao, Zhang, & Liu, 2013). Traces of different engagement levels are integrated in the system of altmetrics, but only a few previous studies have discussed the nuanced view of the types of engagement. From views, saves, discussions, recommendations to citations, there is a natural accession of increasing interest in and level of engagement with the scholarly works (Lin, & Fenner, 2013). These behaviors interact with each other. For example, without viewing the article, discussions on social media and news outlets could hardly take place; at the same time, discussions might in turn induce more views of the article. The views and discussions could both indicate attention from and engagement of readers, but it is hard to separate them when interpreting the impact associated with them. Previous studies have explored how scholarly publications are discussed on social media and news outlets after they are viewed; nevertheless, we take a different approach in this article, by investigating how social media mentions and news reports contribute to the views of articles. In other words, in addition to investigating how people find scholarly articles, we pay particular attention to the role of social media and news outlets in online scholarly communication.

## 2. Methods and Data

*PeerJ* is an open access peer-reviewed journal covering research in the biological and medical sciences. *PeerJ* published its first article in February, 2013, and received its first (partial) Impact Factor of 2.112 in June, 2015. According to their website, *PeerJ* has gained 3.02 million views and 835 thousand downloads of 3,454 articles and preprints, including 1500 peer-reviewed articles and 1561 preprints to the date of December 9, 2015. For each article, *PeerJ* provides article level metrics on its webpage (see Figure 1), including the number of views, downloads, unique visitors, and additionally, referral links that bring 3 or more unique visitors to *PeerJ*.

Data in this study were harvested from *PeerJ*'s website on November 30, 2015. We collected bibliographic metadata and usage data from *PeerJ* website for all the published articles, including DOI, publish date, citations, visitors, views, downloads, as well as Twitter, Facebook, and Google+ activity counts. In addition, we also harvested web referral data of each article, including social referrals and other referrals. Data were parsed and processed into a pre-designed SQL database for the following analyses.

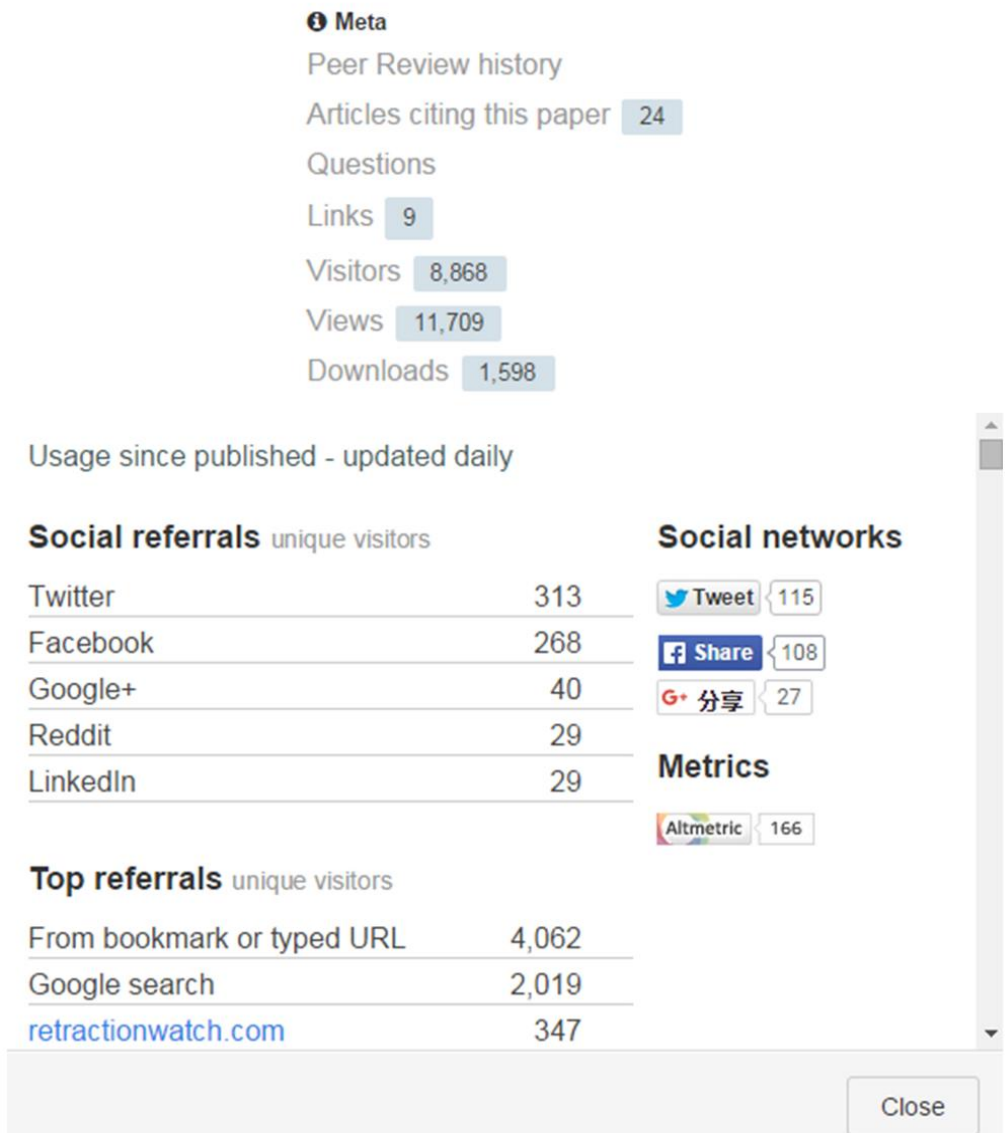

Figure 1 Screenshot of the article level metrics data provided by PeerJ

## 3. Results
*3.1 Statistical analysis*
*3.1.1 Top links to referral sources*

Table 1 lists the top 30 referrals of all the *PeerJ* articles. As Table 1 shows, Google brings the most web traffic to PeerJ articles, including those searching for articles on Google and Google Scholar, plus their different domains in various countries.

Specifically, the first place is the number of unique visitors from bookmark or typed URL, which is about 1.44 million. The number of visitors from Google Search is about 837 thousand, being ranked the second. Other top links are found to come from news reports and blogs (including Io9, Science Daily, and the Annals of Improbable Research, etc.), general search engines (Bing, Yahoo Search and Wikipedia, etc.), academic search engines (National Center for Biotechnolog Information, Web of Science, and DOI.ORG, etc.), as well as social network (Facebook, Twitter and Reddit, etc.).

Table 1 Top 30 referrals links

| rank | referral | visitors | Percent |
|---|---|---|---|
| 1 | From bookmark or typed URL | 1439085 | 42.68% |
| 2 | Google search | 837161 | 24.83% |
| 3 | Twitter | 139323 | 4.13% |
| 4 | m.facebook.com | 96932 | 2.87% |
| 5 | www.ncbi.nlm.nih.gov | 90499 | 2.68% |
| 6 | www.reddit.com | 89754 | 2.66% |
| 7 | www.facebook.com | 81726 | 2.42% |
| 8 | From PeerJ Content Alert Emails | 47601 | 1.41% |
| 9 | io9.com | 10113 | 0.30% |
| 10 | www.bing.com | 9417 | 0.28% |
| 11 | apps.webofknowledge.com | 9190 | 0.27% |
| 12 | www.sciencedaily.com | 7320 | 0.22% |
| 13 | Yahoo search | 7065 | 0.21% |
| 14 | www.improbable.com | 6868 | 0.20% |
| 15 | en.wikipedia.org | 6410 | 0.19% |
| 16 | phenomena.nationalgeographic.com | 6064 | 0.18% |
| 17 | retractionwatch.com | 5841 | 0.17% |
| 18 | www.huffingtonpost.gr | 5328 | 0.16% |
| 19 | dx.doi.org | 5311 | 0.16% |
| 20 | m.reddit.com | 4912 | 0.15% |
| 21 | feedly.com | 4713 | 0.14% |
| 22 | news.sciencemag.org | 4308 | 0.13% |
| 23 | www.lymedisease.org | 4136 | 0.12% |
| 24 | scholar.glgoo.org | 3980 | 0.12% |
| 25 | scikit-image.org | 3730 | 0.11% |
| 26 | www.nature.com | 3584 | 0.11% |
| 27 | www.linkedin.com | 3458 | 0.10% |
| 28 | elpais.com | 3395 | 0.10% |
| 29 | www.theguardian.com | 3313 | 0.10% |
| 30 | www.genscript.com | 3167 | 0.09% |

All the referrals listed in Table 1 were harvested from peerj.com directly and are presented as their original forms. Some referral sources may have different URLs, for example, there are two URLs for Facebook, which are m.facebook.com and www.facebook.com. Some websites have mobile version with the same contents. i.e., m.huffpost.com and huffingtonpost.com. Most referrals in Table 1 are from the United States. Besides those services associated with Google, others include general and scholarly search engines, social networking services, news outlets, and blogs. Yahoo search and Bing are the two search engines following Google. Scholarly search engines include NCBI (National Center for Biotechnology Information) and Web of Science. Housing a series of databases relevant to biotechnology and biomedicine, NCBI focuses on similar disciplines to *PeerJ*'s publication scope.

Facebook and Twitter are two online social networking services founded in the US but widely used all over the world. Slightly different from them, Reddit is also an online bulletin board system (BBS) where community members can submit content and generate discussions. People use these social media for both personal and professional purposes.

In addition, many news websites are also important referrals to *PeerJ*'s scholarly articles: *Io9* is a blog platform focusing on the subjects of science fiction, fantasy, futurism, science, technology and related areas; *Phenomena* is a science salon section of the *National Geographic Journal*; *Science Daily* is an American news website for topical science articles; Feedly is a news aggregator application for various web browsers and mobile devices; news.sciencemag.org is daily news site of the journal *Science*; and *Retraction Watch* is a blog that reports retractions of scientific papers. Most of these news outlets and blogs are focusing on science and technology topics.

Top referrals from other countries, however, are mostly comprehensive news outlets, including huffingtonpost.gr (the Greek domain of the *Huffington Post*, an American online news aggregator and blog), being ranked the 18$^{th}$, *El País* (the highest-circulation daily newspaper in Spain), being ranked the 28$^{th}$, and the *Gardian* (a British news and media website), being ranked the 29$^{th}$. Moreover, nature.com (the website of *Nature* Publishing Group) is ranked the 26$^{th}$. We manually checked the links from nature.com, and found that except for a few exceptions, all of them are from news or blog articles as opposed to scholarly articles. Accordingly, here we classify nature.com as a news website in directing audience to PeerJ.

In this top 30 list, although the counts of visitors from some referrals are large, the numbers of unique websites that contribute to the number of visitors are limited (only 2 or 3). Instances include lymedisease.org (a website about Lyme disease), improbable.com (the website of a scientific humor magazine *Annals of Improbable Research*, which awards Ig Nobel Prize annually), scikit-image.org (an open source image processing library for the Python programming language.), and genscript.com (a company focusing on early drug discovery and development services).

Scholar.glgoo.org, the 24$^{th}$ referral in Table 1, looks similar to Google Scholar. It is one of various mirrors of google.com in mainland China after the domestic banning of Google, circumventing the Great Firewall.

According to our data, the Greek domain of Huffingtonpost brings as many as 5328 visitors to PeerJ. This number is significantly higher than that from huffingtonpost.com. Looking into the original data, we find that all 5328 of the visitors come from one single article (http://www.huffingtonpost.gr/2015/01/15/vythos-plasmata_n_6476472.html). This news introduces PeerJ's article (https://peerj.com/articles/715), studying the size of large sea creatures. Comparing this number to that of visitors directed from the news introducing the same research article on huffingtonpost.com (317 visitors) (http://www.huffingtonpost.com/2015/01/19/sea-creature-sizes-biggest-infographic_n_6487372.html), we believe that this statistic is abnormal. Minor data corruptions were found elsewhere. We paid particular attention to the data used in the last section of our results and manually corrected all the errors that we found.

*3.1.2 Top consolidated referrals*

In fact, many referrals belong to the same second-level domain, i.e., the same company or organization. For example, referrals associated with Google Inc. have hundreds of unique links, including Google the search engine, Google Scholar the academic search engine, as well as diverse specific country domains of them. In addition to that, other products of Google, including Gmail (mail.google.com), Google Drive (drive.google.com), and Google Docs (docs.google.com) are also contributing referrals. Although PeerJ has merged the URLs from the same sources, some URLs are still omitted, e.g., google.com, plus.url.google.com, google.de, etc.

We also found traces of Google Scholar visits by university libraries' patrons (e.g., scholar.google.com.proxy.lib.duke.edu). In these cases, users start Google Scholar search from the library websites, so that they could automatically have direct access to subscription articles already paid for by the university libraries.

As a result, we merge all the subsidiaries of the same company or organization, and present the results in Table 2. About 1.44 million visitors view PeerJ articles from bookmark or typed URL, accounting for 42.68% of the total amount. Google is the second highest referral. As a miscellaneous referrals source, Google brings about 844 thousand visitors, followed by Facebook, Twitter and Reddit, the three largest social networking sites. The next top referral sources include NCBI, From PeerJ Content Alert Emails, Yahoo, Web of Knowledge and Bing.

Table 2 Top 10 consolidated referrals

| rank | referral | visitors | percentage |
|---|---|---|---|
| 1 | From bookmark or typed URL | 1,439,085 | 42.68% |
| 2 | Google | 844,189 | 25.03% |
| 3 | Facebook | 181,434 | 5.38% |
| 4 | Twitter | 139,323 | 4.13% |
| 5 | Reddit | 99,396 | 2.95% |
| 6 | www.ncbi.nlm.nih.gov | 95,162 | 2.82% |
| 7 | From PeerJ Content Alert Emails | 47,601 | 1.41% |
| 8 | Yahoo | 14,043 | 0.42% |
| 9 | Web of Knowledge | 12,901 | 0.38% |
| 10 | Bing | 10,573 | 0.31% |

*3.1.3 Categories*

To gain a deeper understanding of the source of referrals, here we further integrate them into larger categories. There are 12440 unique referral sources to all the *PeerJ* articles that we collected. Some referrals, including "From bookmark or typed URL", Google search, Facebook.com, etc., occur in almost all articles; However, other referrals do not occur as frequently as those mentioned above.

We classify the referrals into nine categories, including:
- General Search Engine
- Academic Resource, including both academic search engines (e.g., Google Scholar, NCBI, Web of Science, etc.), and academic publishers' sites (e.g., *Nature*, *Science*, etc.)
- Social Network

- Bookmark or typed URL
- Mail System
- News & Blog
- RSS
- Institute (university, college, research institutes, or research organization website)
- Others (some technology websites such as Github and Python.org, scientific data repositories such as Dryad, and other referrals that could not be classified into the eight categories above)

For the top 3000 referrals, we manually reviewed the websites and assigned the category accordingly; for the rest, we applied some general rules to determine their categories. For example, google, google.com, yahoo, ask.com were categorized as General Search Engines; while scholar.google.com, apps.webofknowledge.com, and scopus.com were classified as academic engines; referrals containing "mail" were put into the mail category; subdomains of "twitter", "facebook", "linkedin", "google+", "reddit", "weibo", etc., were classified into the Social Network category.

Statistics of the categories are shown in Figure 2. The top category of referrals is Bookmark or typed URL, accounting for almost 43% of all the referrals; General Search Engines bring 879 thousand visitors, accounting for about 26%; News & Blogs and Social Networks are also large categories of referral sources. About 92% of the referrals are from these top four categories. In addition, Academic Resources attract more than 164 thousand visitors, accounting for about 5% of the total counts. Categories of Mail, Institute, RSS and others follow. It is noteworthy that PeerJ includes Google Scholar searches in Google Searches, which means that the share of the Academic Resource category is underestimated. We will analyze details in the following paragraphs.

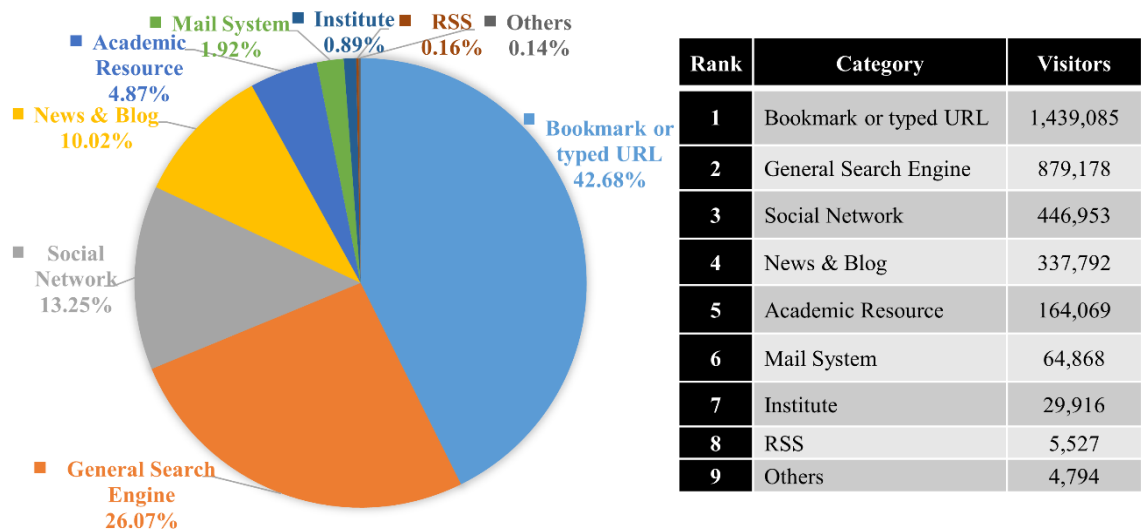

Figure 2 Categories of referrals

Details about the top ten referrals within the above mentioned top four categories are shown in Table 3. Statistics in the categories of Academic Resource, General Search Engine, and Social Network are significantly positively skewed. Despite of the fact that only either one or two major referrals dominate in these categories, here we list all top 10 of them.

In the Academic category, NCBI directs the most visitors to *PeerJ* articles, followed by Web of Knowledge, DOI.ORG, Science and scholar.glgoo.org. It is noteworthy that scholar.glgoo.org, the mirror of Google Scholar, ranks higher than Google Scholar. The reason is that the referral scholar.google.com and its subdomains in different countries are merged into Google search by PeerJ, so that here we only managed to collect the visits from university libraries' proxies. Nevertheless, according to our preliminary data collected in August, 2015, when google.com and Google Scholar were not yet consolidated, visitors from the 95 Google Scholar related referrals contributed 2.91% of the total counts. In other words, if we exclude Google Scholar searches from our General Search Engine category, and add it into the Academic Resource category, the percentages of General Search Engines and Academic Resource should be around 23% and 8%, respectively.

The General Search Engine category is dominated by Google, which refers almost 80 times more people than the second place (Bing), and 100 times more than the third and fourth places (Wikipedia and Yahoo search). We classify Wikipedia into the General Search Engine category, because Wikipedia is most commonly accessed by searching information in general search engines. Then, by clicking on links on the Wikipedia page, audience could be directed to PeerJ. In other words, in the process of scientific information seeking, Wikipedia serves as an intermediate step in the use of general search engines. Among the following referrals, Yandex is from Russia, and Baidu is the largest Chinese search engine. As a localized google substitute of Google in mainland China, where the website of Google can't be accessed, Baidu plays an important role in online searches.

In the Social Network category, Facebook, Twitter and Reddit are the three major referrals, followed by Linkedin, Google Plus, Slashdot, Weibo, Youtube, bit.ly and longurl.org. Sina Weibo (weibo.com) is the most popular Chinese social network service. Since Twitter and Facebook are not accessible in mainland China, Weibo is the main channel via which scholars and the public participate in online social discussion of scientific papers. As for the referrals of bit.ly and longurl.org, which provide URL shortening and expanding services for use in social networking sites, here we classify them into the Social Network category.

In the News & Blog category, the top referrals include science and technology media (Science Daily, the Annals of Improbable, Retraction Watch, Scientific American), mass media (National Geographic, Huffington Post, BBC, Washington Post) and others.

Table 3 Top 10 referrals by different categories

| Rank | Academic Resource | number | General search | number | Social Network | visitors | News & Blog | number |
|---|---|---|---|---|---|---|---|---|
| 1 | NCBI | 95,162 | Google search | 838,199 | Facebook | 181,432 | Io9 | 10,268 |
| 2 | Web of Knowledge | 12,901 | Bing | 10,501 | Twitter | 139,323 | ScienceDaily | 7,324 |
| 3 | dx.doi.org | 7,005 | Wikipedia | 8,290 | Reddit | 99,368 | improbable.com | 6,868 |
| 4 | *Science* | 4,477 | Yahoo search | 8,263 | Linkedin | 3,560 | National Ggeographic | 6,831 |
| 5 | scholar.glgoo.org | 4,120 | ask.com | 1,882 | Google Plus | 3,387 | Huffington Post | 6,616 |
| 6 | *Nature* | 3,663 | Duckduckgo | 1,746 | Slashdot | 3,380 | retractionwatch.com | 5,891 |
| 7 | Scopus | 3,425 | Yandex | 1,492 | Weibo | 2,638 | BBC | 4,749 |
| 8 | Google Scholar | 2,429 | Baidu | 1,410 | youtube.com | 1,851 | lymedisease.org | 4,329 |
| 9 | Wiley | 1,757 | AOL search | 706 | bit.ly | 1,746 | Washington Post | 4,240 |
| 10 | Sciencedirect | 1,605 | myway.com | 625 | longurl.org | 1,383 | Scientific American | 3,870 |

*3.2 From social network mentions to article visiting*

Given the visits data from social media, it is feasible to find out the relationship between social media mentions and their induced article visitor counts. To better understand the role of social media in bringing more readers to academic articles, we conducted a series of correlation analyses. For each article, we filtered out the number of visitors resulted by Facebook and Twitter shares, and looked at if these numbers correlate with the number of mentions on these two platforms. Table 4 summarizes the descriptive statistics and correlation analysis results. Considering that for the most recent articles, it may take time to accumulate the social shares, we excluded articles published after September, 2015. In total, here our final dataset contains 1213 articles.

Since the data are positively skewed, we conducted Spearman correlation analysis. As can be seen in Table 4, social media mentions on Facebook and Twitter are both positively and significantly correlated with the resulted article visitor numbers. In other words, the more social media mentions a PeerJ article receives, the more visitors it has from social media referrals. For Facebook, the correlation coefficient r =0.688 ($p<0.01$); while for Twitter, the coefficient is 0.833 ($p<0.01$).

For the Facebook data, the regression model with the predictor produced $R^2$ = .635, $F(1, 754)$ = 1312.52, $p < .001$. For the Twitter data, the regression model with the predictor produced $R^2$ = .681, $F(1, 1112)$ = 670.43, $p < .001$. As shown in Table 4, for both the Facebook and Twitter data, the social shares have significant positive regression weights, indicating that articles with more mentions are expected to have more visitors.

Table 4 Summary statistics, correlations and results from the regression analysis

| Social media | Variable | N | Mean | Standard deviation | Correlation with visitors | Regression weights | | |
|---|---|---|---|---|---|---|---|---|
| | | | | | | b | β | $R^2$ |
| Facebook | visitors | 943 | 102.87 | 694.666 | | | | |
| | shares | 871 | 27.62 | 91.815 | .688** | 6.301*** | .174 | .635 |
| Tweet | visitors | 1114 | 62.62 | 260.930 | | | | |
| | shares | 1213 | 14.98 | 41.107 | .833** | 3.751*** | .145 | .376 |

Note: * $p < .05$   ** $p < .01$   ***$p < .001$

*3.3 From news outlets to article visiting*

What about the relationship between news mentions and their induced article visitor counts? To explore this question, we look at the referrals of two scholarly journals and several mainstream media, and the visitors they bring to PeerJ articles. Considering the significantly lower number of articles that have been reported in news outlets, correlation analysis is not applicable. However, we could still look at the descriptive statistics to get a sense of the effects of news reports on visitor counts. To make sure that the sources are news articles as opposed to blog articles or scholarly articles, we manually checked all the articles and filtered out news articles.

In our dataset, 7 major news media report 41 PeerJ papers: 16 reported by *The Huffington Post*, 15 reported by BBC, 12 reported by *The Washington Post* and *The Guardian*, 5 reported by

NPR, *Wired* Magazine, and *The New York Times*. The effects of news reports on articles' visitor counts are diverse. In other words, there may exist vast differentiation for the caused visits when scientific papers are reported by news media; and this differentiation is dependent not only on the what the article is about, but also on the what the media are. As is shown in Table 5, article No. 8 (https://peerj.com/articles/8) gets 83 visits from Huffington Post; while article No. 338 (https://peerj.com/articles/338) gets 61 visits from BBC, 1000 visits from The Guardian, 1999 visits from Washington Post, but only 22 visits from Huffington Post. The cells in Table 5 are shaded with gradations of colors from light blue (minimum value) to dark blue (maximum value).

Table 5 Number of article visitors caused by news report

| Article number | BBC | The Guardian | Washington Post | Huffington Post | NPR | New York Times | Wired |
|---|---|---|---|---|---|---|---|
| 182 | | 138 | | 9 | | 162 | |
| 338 | 61 | 1100 | 1999 | 22 | | | |
| 857 | 232 | 483 | 51 | 46 | 89 | | |
| 885 | 572 | 133 | 26 | | | | |
| 1011 | 498 | 194 | 125 | 16 | | | |
| 1258 | 1254 | | 489 | | 1452 | | 1821 |
| 8 | | | | 83 | | | |
| 36 | | 81 | | 2 | | | |
| 60 | | | | 15 | | | |
| 62 | 161 | | | | | | |
| 103 | 283 | | | | 85 | | 86 |
| 156 | | | | 82 | | | |
| 175 | | 49 | | | | | |
| 209 | | 165 | | | | | |
| 218 | 312 | | | | | | |
| 278 | | | | | | 562 | 138 |
| 283 | | | | | | | 100 |
| 297 | | | | | 29 | | |
| 318 | | | | 24 | | 92 | |
| 424 | 97 | | | 13 | | | |
| 445 | | | | | | 129 | |
| 447 | | | | 25 | | | |
| 471 | 141 | | | | | | |
| 488 | 30 | | | | | | |
| 519 | | | 56 | | | | |
| 523 | | | | 139 | | | |
| 556 | | | | | | 442 | |
| 570 | 72 | | | | 128 | | |
| 574 | | | 19 | | | | |
| 579 | | | | | | | 118 |
| 652 | | | | 30 | | | |
| 715 | | | 755 | 347 | | | |
| 798 | | 339 | | | | | |
| 830 | | | | 163 | | | |
| 854 | | 304 | | | | | |
| 1037 | | | 302 | 207 | | | |
| 1053 | 290 | | | | | | |
| 1124 | 661 | 146 | | 48 | | | |
| 1140 | | 172 | 300 | | | | |
| 1155 | 84 | | | | | | |
| 1227 | | | | 56 | | | |

## 4 Discussion

As we mentioned in the Introduction section, it is the information behaviors associated with the digital traces that could provide more insights into how the scholarly information is being used.

We find that a fairly large portion of the article visits are directed from General Search Engines, including Google, Bing and Yahoo, etc. No matter what search queries were used, and no matter what the purposes of the search were, the visitors found these PeerJ articles by searching on general search engines. This indicates that general search engines are commonly used tools to find scholarly articles. Aided by the open movements, general search engines indexing information of scholarly articles are making it easier for both scholars and the public to find scholarly information online. However, without more detailed demographic information, it is difficult to determine whether the audience are scholars or the public. It would be arbitrary to conclude that scholars are inclined to use academic search engines while the public are more likely to use general search engines to search for scholarly communication. As a result, we need to analyze searching behaviors separately.

Without further research, we cannot tell whether people prefer to use general search engines or not; however, it is true that under certain circumstances, it is more efficient and effective to use general search engines as opposed to scholarly search engines, even for researchers who are familiar with academic databases. For instance, using handy general search engine could help return the best answer in a focalized search. When they remember piece of information about an existing article (say, keywords in the title, one author's name, or even only the data explored), or when they find an interesting article mentioned on a colleague's tweet but with no links provided, a simple search on general search engine could lead them to the answer efficiently. Search engines integrated into browsers makes it even more convenient to use general search engines – by typing in the address bar and press Enter, users could even skip the step of opening the search engine page. More importantly, some scholarly search engines are behind pay walls, so that having to firstly logging in either makes the search more difficult or impossible. Nevertheless, using academic search engine would be beneficial in exploratory search when scholars would like to retrieve extensive scholarly information about one topic. Under this circumstance, a general search engine usually returns too much information with a lower precision, while an academic search engine could provide a higher precision and at the same time guarantee a high recall.

The public encompass a wider range of audience, including students, scientific practitioners, and people from any professions. Their purposes of searches might not be finding a scholarly article to read at the first place, but they somehow find the scholarly articles of interest or value to them. While students and educators might read scholarly articles for educational purposes; housewives are interested in finding scientific guidelines to provide healthier meals for their families; and beginning investors might be willing to learn some basic financial knowledge and skills by reading scholarly articles, etc. As more people are getting highly educated, seeking and utilizing scholarly information are no longer limited to scholars. In fact, it is beneficial for both scholars and the public to get scholarly information widely spread and sufficiently understood.

Besides general and academic search engines, the top referrals are either social networking platforms, or news outlets and blogs. Similar to the search behaviors, it is difficult to distinguish scholars' activities from those of the public. More demographic data could help in analyzing who the people are, but still, more qualitative studies need to be completed to better understand why and how the articles are introduced, discussed or recommended in the social networks,news and blogs.

When studying the online communication of scholarly articles, many previous studies try to use Twitter mentions, news reports, and blog recommendations as proxies to evaluate the articles' impact. However, no previous studies have explored whether and how these activities could facilitate the dissemination of impacts in detail. There is a long way to go from having the scholarly information at sight to being influenced by the information; how strong the impact would be also depends on various factors. For instance, a tweet could be posted with no one seeing it; someone could skim that tweet, without clicking on the link to the original article; even if someone visited the original scholarly article, he/she could simply read the title or abstract; etc. Most of these behaviors could hardly be tracked on a large scale, due to technical and privacy limitations; accordingly, it is difficult to study whether the article has truly impacted the audience or not. To study the channel and strength of the impact flow would be even more challenging. However, our quantitative empirical study has demonstrated that social network discussions and news reports do attract more visitors to scholarly articles. Although visitors don't necessarily mean readers, visiting scholarly articles is already one step further than reading only the social media and news media content mentioning the scholarly articles. We could safely conclude that social media and news media not only spread a basic sense of scholarly articles, but also bring some visitors to the articles themselves. This brings potential opportunities to study why these visitors visit the articles, and how they are influenced by the articles.

## Acknowledgments

The work was supported by the project of "National Natural Science Foundation of China" (61301227), the project of "Growth Plan of Distinguished Young Scholar in Liaoning Province" (WJQ2014009), and the project of "the Fundamental Research Funds for the Central Universities" (DUT15YQ111).